\documentclass{elsart}
\usepackage{epsfig}
\begin{document}
\title{Application of nonextensive statistics to particle
and nuclear physics}
\author{G. Wilk$^{a}$, Z.W\l odarczyk$^{b}$}
\address{$^{a}$The Andrzej So\l tan Institute for
 Nuclear Studies\\Ho\.za 69; 00-689 Warsaw, Poland; e-mail:
 wilk@fuw.edu.pl\\
$^{b}$Institute of Physics, Pedagogical University\\
Konopnickiej 15; 25-405 Kielce, Poland; e-mail: wlod@pu.kielce.pl\\
 \today}
{\scriptsize Abstract: We present an overview of possible imprints of
non-extensitivity in particle and nucler physics. Special emphasis is
put on the intrinsic fluctuations present in the system under
consideration as the possible source of nonextensivity. The possible
connection of nonextensivity and the self organized criticality
apparently being observed in some cosmic rays and hadronic
experiments will also be discussed. }
\section{Introduction - how we got to it}
Our encounter with notion of nonextensivity originated with
observation that in some cosmic ray data (like depth distribution of
starting points of cascades in Pamir lead chamber \cite{WWCR}) one
encounters deviations from the usual exponential distributions of
some variables towards the power-like ones:
\begin{equation}
\frac{dN}{dT}\, =\, {\rm const}\cdot \exp \left(\, -\,
\frac{T}{\lambda}\, \right) \Rightarrow
{\rm const}\cdot \left[1\, -\,
      (1-q)\frac{T}{\lambda}\right]^{\frac{1}{1-q}}. \label{eq:EXP}
\end{equation}
Here $N$ denotes the number of counts at depth $T$ (cf. \cite{WWCR}
for details). Whereas in \cite{WWCR} we have proposed as explanation
a possible fluctuations of the mean free path $\lambda$ in eq.
(\ref{eq:EXP}) characterised by relative variance $\omega\, =\,
\frac{\left(\langle \sigma^2\rangle - \langle \sigma \rangle
^2\right)}{\langle \sigma \rangle ^2}\, \geq\, 0.2 $,
in \cite{CR} the same data were fitted by power-like (L\'evy type)
formula as above keeping $\lambda$ fixed and setting $q=1.3$. In this
way we have learned about Tsallis statistics and Tsallis nonextensive
entropy and distributions \cite{T}. Following this approach we have
then demonstrated \cite{WW,DENTON} that, indeed,
\begin{equation}
L = \exp\left(-\frac{x}{\lambda_0}\right)~~ \Rightarrow~~
L_q = \exp_q\left(-\frac{x}{\lambda_0}\right) =
\left\langle \exp\left(-\frac{x}{\lambda}\right)\right\rangle ,
\label{eq:Nonext}
\end{equation}
with $q = 1 \, \pm \, \omega $ for $q>1~(+)$ and $q<1~(-)$, i.e.,
there is connection between the measure of fluctuations $\omega$ and
the measure of nonextensivity $q$ (it has been confirmed recently in
\cite{BECKA}). This will be the first subject discussed here. The
other will be connected with the self-organized criticality form of
nonextensivity as apparently seen in hadronic processes.

\section{Temperature fluctuations?}

The most interesting example of the possible existence of such
fluctuations is the trace of power like behaviour of the
transverse momentum distribution in multiparticle production
processes encountered in heavy ion collisions \cite{ALQ,UWW}. Such
collisions are of special interest because they are the only place
where new state of matter, the Quark Gluon Plasma, can be produced
\cite{QGP}. Transverse momentum distributions are believed to provide
information on the temperature $T$ of reaction, which is given by the
inverse slope of $dN/dp_T$, if it is exponential one. If it is not,
question arises what we are really measuring. One explanation is the
possible flow of the matter, the other, which we shall follow here,
is the nonextensivity (or rather fluctuations leading to it).
Namely, as was discussed in detail in \cite{ALQ} the extreme
conditions of high density and temperature occuring in
ultrarelativistic heavy ion collisions can invalidate the usual BG
approach and lead to $q>1$, i.e., to
\begin{equation}
\frac{dN(p_T)}{dp_T} = {\rm const}\cdot \left[ 1\, - (1 - q)
\frac{\sqrt{m^2 + p^2_T}}{T}\right]^{\frac{1}{1-q}}. \label{eq:Pt}
\end{equation}
Here $m$ is the mass of produced particle and $T$ is, for the $q=1$
case, the {\it temperature} of the hadronic system produced. Although
very small ($|q-1| \sim 0.015$) this deviation, if interpreted
according to eq. (\ref{eq:Nonext})), leads to quite large relative
fluctuations of temperature existing in the nuclear collisions,
$\Delta T/T \simeq 0.12$. It is important to stress that these are
fluctuations existing in small parts of hadronic system in respect to
the whole system rather than of the event-by-event type for which,
$\Delta T/T = 0.06/\sqrt{N} \rightarrow 0$ for large $N$ (cf.
\cite{OFT} for relevant references). Such fluctuations are
potentially very interesting because they provide direct measure of
the total heat capacity $C$ of the system:
\begin{equation}
\frac{\sigma^2(\beta)}{\langle \beta\rangle ^2}\, =\, \frac{1}{C}\,
=\, \omega\, =\, q - 1 \label{eq:C}
\end{equation}
($\beta =\frac{1}{T}$) in terms of $\omega = q - 1$. Therefore,
measuring {\it both} the temperature of reaction $T$ and (via
nonextensivity $q\neq 1$) its total heat capacity $C$, one can not
only check whether an approximate thermodynamics state is formed in a
single collision but also what are its theromdynamical properties
(especially in what concerns the existence and type of the possible
phase transitions \cite{OFT}).

To observe such fluctuations an event-by-event analysis of data is
needed \cite{OFT}. Two scenarios must be checked: $(a)$ $T$ is
constant in each event but because of different initial conditions it
fluctuates from event to event and $(b)$ $T$ fluctuates in each event
around some mean value $T_0$. Fig. 1 shows typical event obtained in
simulations performed for central $Pb+Pb$ collisions  taking place
for beam energy equal $E_{beam}=3~A\cdot$TeV in which density of
particles in central region (defined by rapidity window $-1.5 <y<
1.5$) is equal to $\frac{dN}{dy} = 6000$ (this is the usual value
given by commonly used event generators \cite{MC}). In case $(a)$ in
each event one expects exponential dependence with $T=T_{event}$ and
possible departure from it would occur only after averaging over
all events. It would reflect fluctuations originating from different
initial conditions for each particular collision. This situation is
illustrated in Fig. 1a where $p_T$ distributions for $ T = 200$ MeV
(black symbols) and $T = 250$ MeV (open symbols) are presented. Such
values of $T$ correspond to typical uncertainties in $T$ expected at
LHC accelerator at CERN. Notice that both curves presented here are
straight lines. In case $(b)$ one should observe departure from the
exponential behaviour already on the single event level and it
should be fully given by $q>1$. It reflects situation when, due to
some intrinsically dynamical reasons, different parts of a given
event can have different temperatures \cite{WW,DENTON}. In Fig. 1b
black symbols represent exponential dependence obtained for $T = 200$
MeV (the same as in Fig. 1a), open symbols show the power-like
dependence as given by (\ref{eq:Pt}) with the same $T$ and with
$q=1.05$ (notice that the corresponding curve bends slightly upward
here). In this typical event we have $\sim 18000$ secondaries, i.e.,
practically the maximal possible number. Notice that points with
highest $p_T$ correspond already to single particles. As one can see,
experimental differentiation between these two scenarios will be very
difficult, although not totally impossible. On the other hand, if
successful it would be very rewarding - as we have stressed before.

One should mention at this point that to the same cathegory of
fluctuating temperature belongs also attempt \cite{BCM} to fit energy
spectra in both the longitudinal and transverse momenta of particles
produced in the $e^+e^-$ annihilation processes at high energies,
novel nonextensive formulation of Hagedorn statistical model of
hadronization process \cite{Beck,BECKA} and description of single
particle spectra \cite{UWWF}.

\section{Nonexponential decays}

Another hint for intrinsic fluctuations operating in the physical
system could be the known phenomenon of nonexponential decays
\cite{NEXPDEC}. Spontaneous decays of quantum-mechanical unstable
systems cannot be described by the pure exponential law (neither for
short nor for long times) and survival time probability is
$P(t)\propto t^{- \delta}$ instead of exponential one. It turns out
\cite{NEXPDEC} that by using random matrix approach, such decays can
emerge in a natural way from the possible fluctuations of parameter
$\gamma = 1/\tau$ in the exponential distribution $P(t) =
\exp(-\gamma t)$. Namely, in the case of multichannel decays (with
$\nu$ channels of equal widths involved) one gets fluctuating widths
distributed according to gamma function
\begin{equation}
P_{\nu}(\gamma) = \frac{1}{\Gamma\left(\frac{\nu}{2}\right)}
                    \left(\frac{\nu}{2<\gamma>}\right)
           \left(\frac{\nu \gamma}{2<\gamma>}\right)^{\frac{\nu}{2}-1}
           \exp\left(- \frac{\nu \gamma}{2<\gamma>}\right)
\label{eq:FINALG}
\end{equation}
and strength of their fluctuations is given by relative variance
$\frac{\left\langle (\gamma \, -\,
<\gamma>)^2\right\rangle}{<\gamma>^2}\, = \, \frac{2}{\nu}$, which
decreases with increasing $\nu$.
According to \cite{WW}, it means therefore that, indeed,
\begin{equation}
L_q(t,\tau_0) = \frac{2-q}{\tau_0}\, \left[1\, -\,
(1-q)\frac{t}{\tau_0}\right]^{\frac{1}{1-q}} , \label{eq:Pq}
\end{equation}
with the nonextensivity parameter equal to $q = 1 + \frac{2}{\nu}$.

\section{Self-organized criticality in cascade processes}

For non-equilibrium phenomena the sources of power-law distributions
are self-organized criticality (SOC - nonequilibrium systems are
continuously driven by their own internal dynamic to a critical state
with power-laws omnipresent) \cite{SOC} and stochastic multiplicative
processes (power-law is generated by the presence of underlying
replication of events) \cite{SMP}. All they can be unified in terms
of generalized nonextensive statistics \cite{T}. It has been recently
argued \cite{MENG,JMRT} that in systems of quarks and gluons formed
in collision processes there exists evidence of SOC behaviour. The
picture proposed in \cite{MENG} is that system of interacting soft
gluons can be, and should be, considered as an open, dynamical,
complex system with many degrees of freedom, remaining in general far
from thermal and/or chemical equilibrium. Applied first to
description of formation of colour-singlet gluon clusters in
inelastic diffraction scattering this argumentation has been recently
extended to cover also quarks \cite{JMRT}. This allowed to describe
the existing data on high transverse momentum jet production in a
uniform way (jet is, loosely speaking, a bunch of collimated hadrons
going in one direction). Namely, the $E_T$ jet cross section ($E_T$
is the energy  of such jet measured in direction perpendicular to the
collision axis) is given by simple power law:
\begin{equation}
\frac{d^2\sigma}{dE_T d\eta} \propto E_T^{-\alpha} , \label{eq:MENG}
\end{equation}
with $\alpha \sim 5,~7$ and $9$ for small, intermediate and large
values of $E_T$, respectively, indicating three different possible
scenarios operating at different ranges of $E_T$ (according to
\cite{JMRT}).

Similar situation has been encountered in cosmic ray physics
where energy spectra of particles from atmospheric cascades
exhibit power-like behaviour, which can be described by L\'evy type
distribution \cite{RWW}. For sufficiently large energy fraction $x_N$
allocated to the cascade with $N$ generations it is given by
\begin{equation}
P(x_N) \, \propto\, \left(\frac{x_N}{\langle
x_N\rangle}\right)^{-\alpha}, \qquad \alpha=\frac{1}{q-1} ,
\label{eq:INF}
\end{equation}
where $q$ is nonextensivity index depending on the number of
generations $N$, $q\, =\, \frac{3}{2}\, -\, \frac{c^{N-1}}{2}$,
($c$ is $N$-independent parameter which should be fitted, here
$c=0.55$). For events without cascading $N=1$ and $q=1$, i.e., one
gets usual exponential distribution. For large $N$ (in practice for
$N\ge 6$) $q\rightarrow 3/2$ (and $\alpha \rightarrow 2$ in
(\ref{eq:INF})), which is limiting value available here for $q$
(explaining why in such experiments one always observes $\alpha \le
2$). This result tells us that the exactly power-like behaviour of
spectra is achieved only asymptotically, for long enough cascades.
This has been clearly demonstrated in \cite{RWW} on some experimental
data.

Coming back to eq. (\ref{eq:MENG}) we argue that, as can be seen in
Fig. 2, the same data can be described by a suitably modified
equivalent of eq. (\ref{eq:INF}):
\begin{equation}
\frac{d^2\sigma}{dE_T d\eta}\vert_{\eta=0} =
      c\cdot E_T^{-\alpha}\cdot[1 - {\rm Erf}(a)], \label{eq:NEWMENG}
\end{equation}
this time with only one exponent for all values of $E_T$: $\alpha=
5.01$ and $4.9$ for energies $\sqrt{s}=546$ and $1800$ GeV considered
in \cite{JMRT}, respectively. The correction factor to the power law
comes from accounting for smearing out of the initial conditions of
the cascade (not present in (\ref{eq:INF})) with $a =
(1-\alpha)(\frac{\delta}{\sqrt{2}}) + \frac{\sqrt{2}}{\delta}\ln
\Delta_0$ and $\Delta_0 = \frac{E_T}{E_T^{(0)}}$ (ratios of the
actual energy of jet $E_T$ to its threshold energy  $E_T^{(0)}$).
This smearing was assumed in the following log-normal form:
$P(\Delta)d(\ln\Delta) = \frac{1}{\sqrt{2}\delta}\exp\left[-
\frac{\left(\ln \Delta - \ln \Delta_0\right)^2}{2\delta^2}\right]$.
Parameter $\delta$ is dispersion of the smearing distribution and is
set to be equal $\delta = 0.74$ and $0.79$ for the above respective
energies. The value $\alpha \sim 5$ in (\ref{eq:NEWMENG}) emerges in
a natural way from that of $2$ in (\ref{eq:INF}). The point is that
the main interaction in hadronic collisions proceeds between quarks,
which are produced in the quark-gluon QCD cascade process from the
initial dressed valence quarks in nucleon. Therefore jets are
produced by collision of two such quarks with energy fractions $x$
and $y$ and energy $M$ of such collision is $M = \sqrt{z\cdot s} \sim
\sqrt{z}$ (where $z=x\cdot y$ and where $s$ is initial invariant
energy of reaction squared). Therefore distribution in
$z$ is of the form of convolution: $P(z) = P(x)\otimes P(y) \sim
z^{-3}$, which results in 
\begin{equation}
\frac{d\sigma}{dz}\, =\, \frac{1}{M}\frac{d\sigma}{dM} \sim
M^{-6}\qquad {\rm  or}\qquad \frac{d\sigma}{dM}\, \sim\, M^{-5}.
\label{eq:M5}
\end{equation}
Because $E_T\sim M$ eq.(\ref{eq:M5}) becomes eq.(\ref{eq:MENG}).
Therefore value of $\alpha = 5$ is what really comes from the pure
cascade. All deviations from it needed to fit data come from the
initial conditions properly accounted for.

\section{Summary}

There is steadily growing evidence that some peculiar features
observed in particle and nuclear physics (including cosmic
rays) can be most consistently explained in terms of the suitable
applications of nonextensive statistic of Tsallis. Here we were able
to show only some selected examples, more can be found in
\cite{DENTON}. However, there is also some resistance towards
this idea, the best example of which is provided in \cite{GG}. It is
shown there that mean multiplicity of neutral mesons produced in
$p-\bar{p}$ collisions as a function of their mass (in the
range from $m_{\eta} = 0.55$ GeV to  $M_{\Upsilon}=9.5$ GeV) and
the transverse mass $m_T$ spectra of pions (in the range of $m_T
\simeq1\div15$ GeV), both show a remarkable universal behaviour
following over $10$ orders of magnitude the same {\it power law}
function $C\cdot x^{-P}$ (with $x=m$ or $x=m_T$) with $P\simeq 10.1$
and $P\simeq 9.6$, respectively. In this work such a form was just
{\it postulated} whereas it emerges naturally in $q$-statistics
with $q= 1 + 1/P \sim 1.1$ (quite close to results of \cite{BCM}). We
regard it as new, strong imprint of nonextensivity present in
multiparticle production processes. This interpretation is
additionally supported by the fact that in both cases considered in
\cite{GG} constant $c$ is the same. Apparently there is no such
phenomenon in $AA$ collisions which has simple and interesting
explanation: in nuclear collisions volume of interaction is much
bigger what makes the heat capacity $C$ also bigger. This in turn,
cf. eq.(\ref{eq:C}), makes $q$ smaller. On should then, indeed, expect
that $q_{hadronic} >> q_{nuclear}$, as observed.

%\newpage

\noindent
{\bf Acknowledgments:} GW is gratefull to the organizers of the {\it
NEXT2001} for their support and hospitality. The partial support of
Polish Committee for Scientific Research (grants 2P03B 011 18 and
621/E-78/SPUB/CERN/P-03/DZ4/99) is acknowledged. \\

%\newpage

\begin{figure}[ht]
  \begin{minipage}[ht]{65mm}
    \centerline{
        \epsfig{file=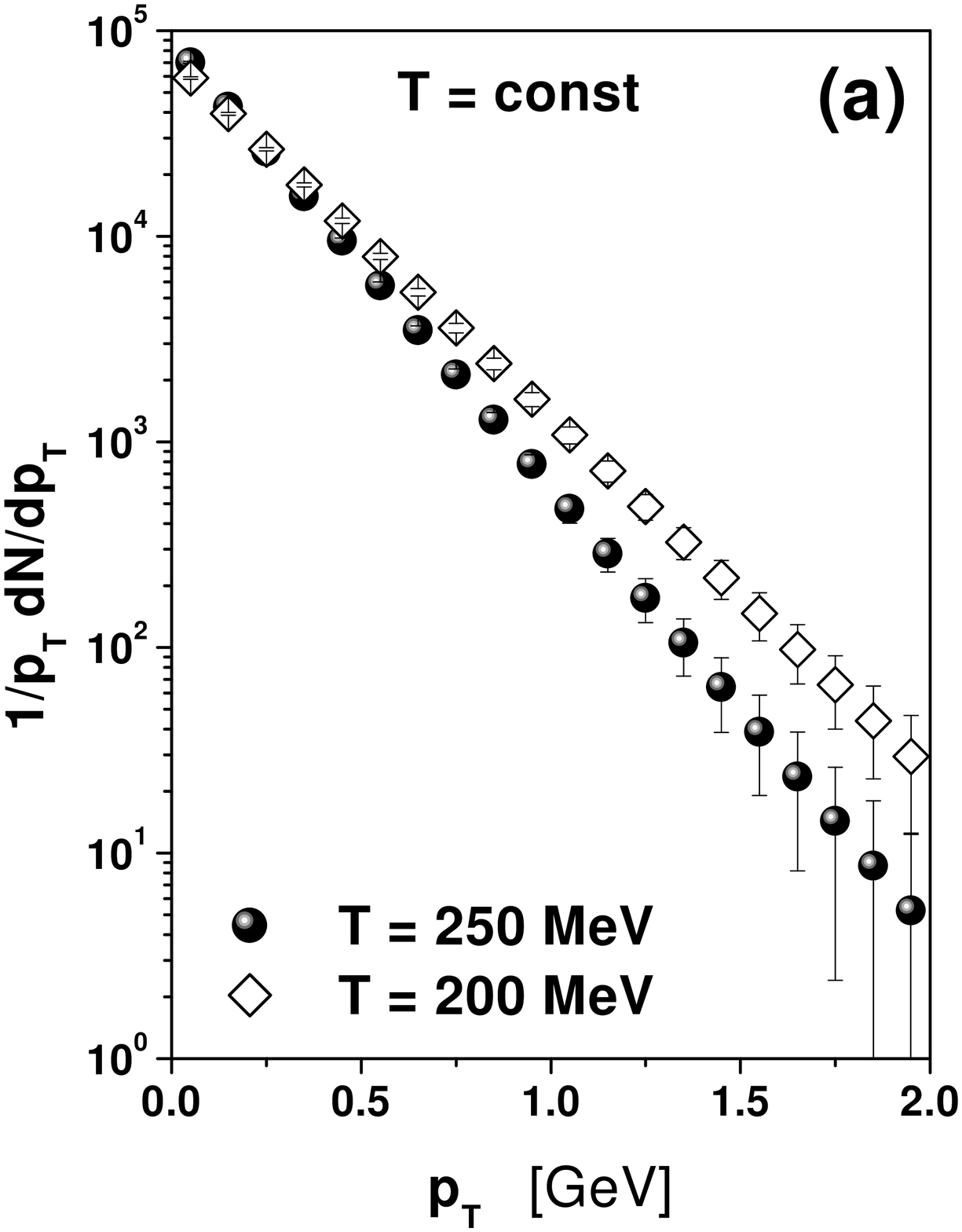, width=60mm}
     }
  \end{minipage}
\hfill
  \begin{minipage}[ht]{65mm}
    \centerline{
       \epsfig{file=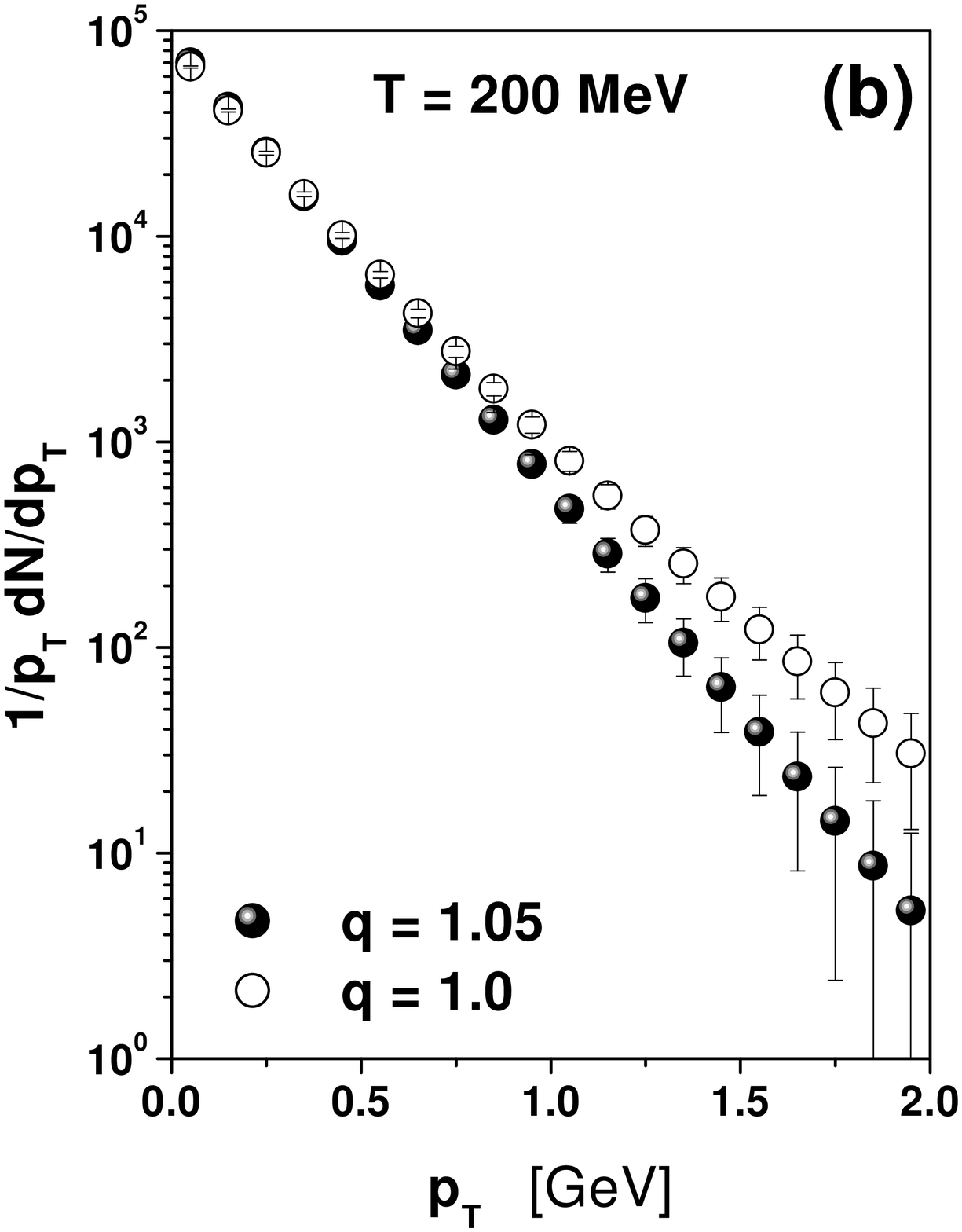, width=60mm}
     }
  \end{minipage}
  \caption{\footnotesize $(a)$ Normal exponential $p_T$ distributions
                     (i.e., $q=1$) for $T = 200$ MeV (black symbols)
                     and $T = 250$ MeV open symbols). $(b)$ Typical
                     event from central $Pb+Pb$ at
                     $E_{beam}=3~A\cdot$TeV (cf. text for other details)
                     for $=200$ MeV for $q=1$ (black symbols) exponential
                     dependence and $q=1.05$ (open symbols).
}
  \label{Figure1}
\end{figure}

\begin{figure}[ht]
  \begin{minipage}[ht]{65mm}
    \centerline{
        \epsfig{file=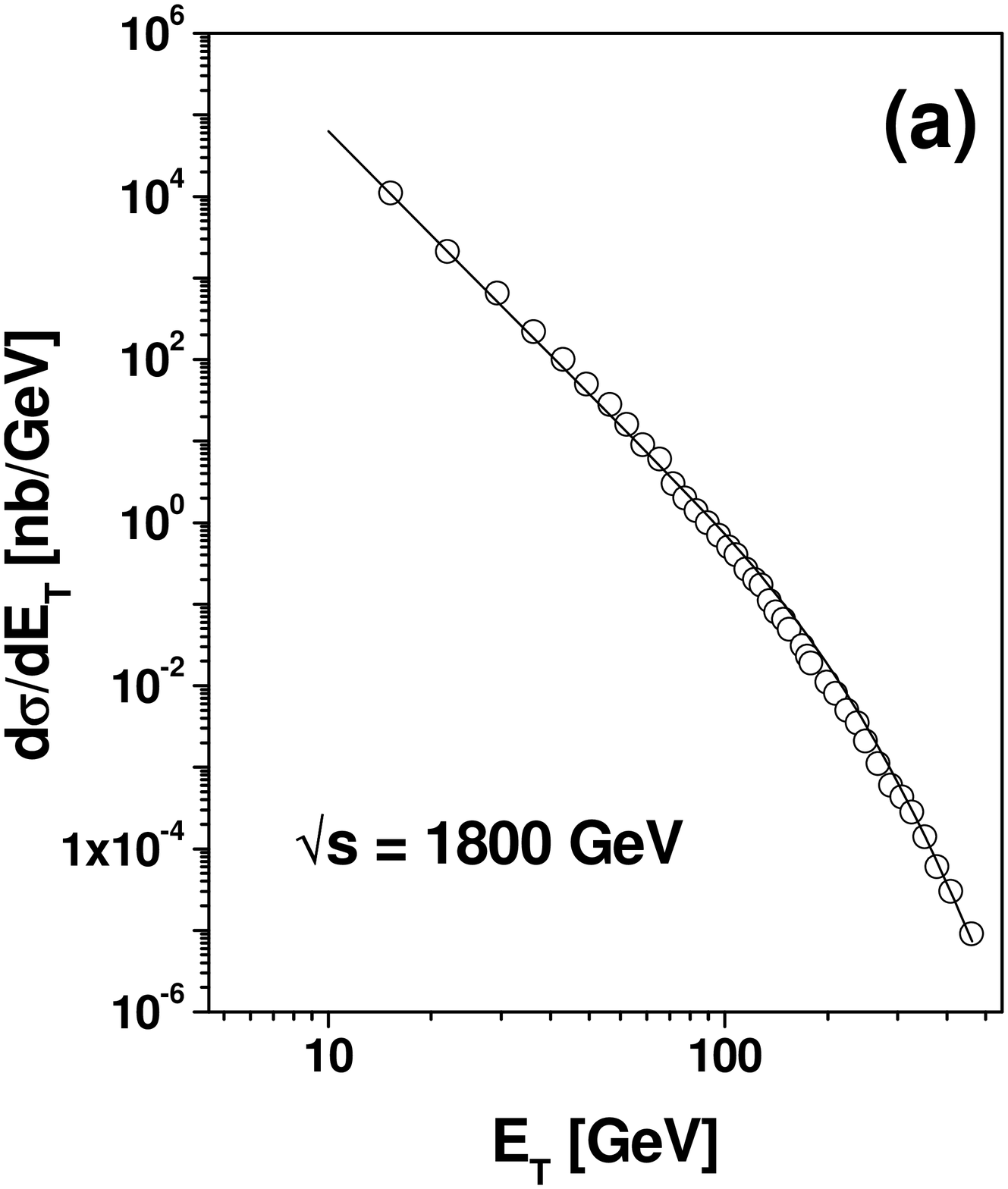, width=60mm}
     }
  \end{minipage}
\hfill
  \begin{minipage}[ht]{65mm}
    \centerline{
       \epsfig{file=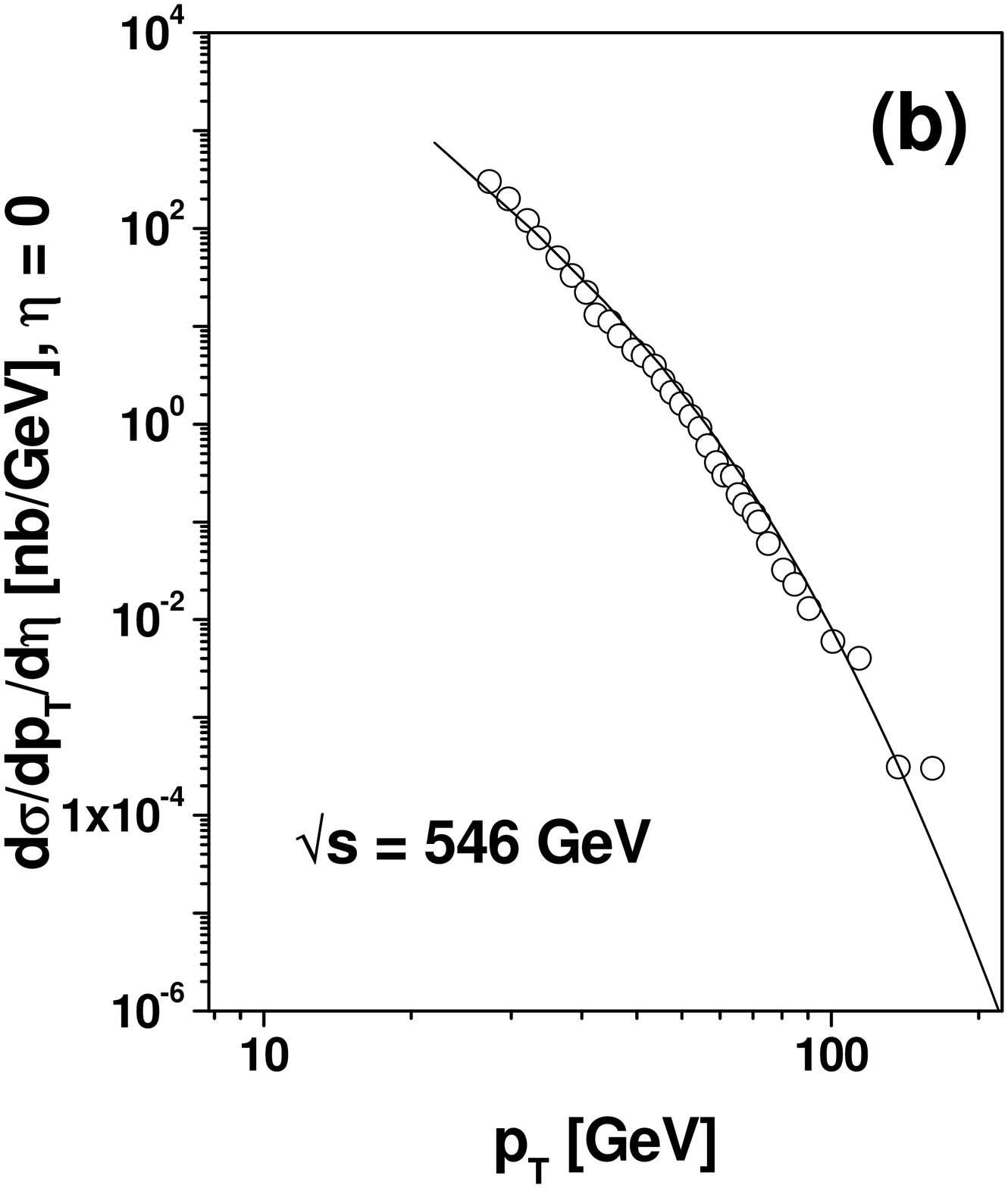, width=60mm}
     }
  \end{minipage}
  \caption{\footnotesize The high energy $\bar{p}p$ collisions
data used
                     in \cite{JMRT} (cf. their Fig. 1) fitted by our
                     eq. (\ref{eq:NEWMENG}). See text for details.
}
  \label{Figure2}
\end{figure}

\end{document}